\documentclass[twoside]{revtex4}
\usepackage{graphicx}
\usepackage{fancyhdr}
\usepackage{epsfig}
\usepackage{subfigure}
\def\beq{\begin{equation}}
\def\eeq{\end{equation}}
\def\bea{\begin{eqnarray}}
\def\eea{\end{eqnarray}}
\begin{document}

\title{Holographic QCD predictions for rare B decays}

%

\author{Mohammad Ahmady\footnote{Speaker}}
\affiliation{Department of Physics, Mount Allison University, Sackville, New Brunswick E4L 1E6, Canada}
\author{Ruben Sandapen}
\affiliation{Department of Physics, Acadia University, Wolfville, Nova Scotia B4P 2R6, Canada}

\begin{abstract}
Light-front wavefunctions obtained from holographic AdS/QCD are used to obtain the distributions amplitudes for light mesons. Consequently, alternate predictions for B transition to light mesons form factors are presented. In this talk, I compare our results for rare B decays to those obtained from QCD sum rules and available experimental data. 
\end{abstract}

\maketitle

\thispagestyle{fancy}


\section{Introduction}
Rare flavour changing neutral current  B-meson decays have been under theoretical and experimental scrutiny in the past three decades.  B-factories and other dedicated experimental facilities have provided data with ever increasing precision on more variety of rare decay channels.  The inclusive decays, which are easier to handle theoretically, are extremely challenging for measurement especially in a hadronic environment like LHCb.  Exclusive modes, on the other hand, suffer from hadronic uncertainties and therefore, the better understanding of the non-perturbative QCD effects is essential in comparing the standard model (SM) predictions against the experimental data.  This contribution focuses on the $B \to K^*$ transition but is more broadly based on Refs. \cite{Ahmady:2018fvo,Ahmady:2015fha,Ahmady:2014sva,Ahmady:2014cpa,Ahmady:2013cva}. 

Predicting observables in rare B decays requires a knowledge of non-perturbative quantities like transition form factors as well as decay constants. These quantities are related to the meson Distribution Amplitudes (DAs). Traditionally, DAs are obtained using QCD Sum Rules. Here, we derive them using the holographic light-front meson wavefunction which is itself obtained by solving the holographic Schr\"odinger Equation for mesons. The differences between our predictions and those of the QCD sum rules (QCDSR) offers an opportunity to gauge the impact of hadronic uncertainties in the theory predictions of observables in rare B decays. This can be helpful to filter out genuine New Physics signals in these decays.

\section{Holographic light-front QCD}
In semiclassical light-front QCD, where quark masses and quantum loops are neglected, the transverse part of the valence light-front wavefunction of a meson of mass $M$, satisfies the so-called holographic Schr\"odinger Equation \cite{deTeramond:2008ht}
\begin{equation}
\left(-\frac{\mathrm{d}^2}{\mathrm{d}\zeta^2}-\frac{1-4L^2}{4\zeta^2} + U_{\mathrm{eff}}(\zeta) \right) \phi(\zeta)=M^2 \phi(\zeta) \;,
\label{hSE}
\end{equation}
where $\mathbf{\zeta} = \sqrt{z\bar{z}} b$ ($\bar{z} = 1-z$). $b$ is the transverse separation of the quark and antiquark and $z$ is the light-front momentum fraction carried by the quark. Eq. (\ref{hSE}) is called holographic because it maps onto the wave equation for string modes propagating in the higher $5$-dimensional anti-de Sitter spacetime. This holographic mapping, together with the requirement to introduce a mass scale in Eq. (\ref{hSE}) while preserving the conformal invariance of its underlying action, leads to a unique form for the confining potential:\begin{equation}
U_{eff}(\zeta)=\kappa^4 \zeta^2 + 2 \kappa^2 (J-1)\;.
\end{equation}
The holographic Schr\"odinger Equation can then be solved:
\begin{equation}
M^2=(4n+2L+2)\kappa^2 + 2\kappa^2(J-1)=4\kappa^2 (n+L+\frac{S}{2})\;,
\label{mass}
\end{equation}
\begin{equation}
\phi_{n,L}(\zeta)= \kappa^{1+L} \sqrt{\frac{2 n !}{(n+L)!}} \zeta^{1/2+L} \exp{\left(\frac{-\kappa^2 \zeta^2}{2}\right)} L_n^L(z^2 \zeta^2) \;,
\label{phi-zeta}
\end{equation}
where $n$, $L$ and $S$ are the principal, orbital and spin quantum numbers respectively. As can be seen from Eq. (\ref{mass}), the Regge slope for vector mesons determines the fundamental scale of the model:  $\kappa =0.54$ GeV \cite{Brodsky:2014yha}. The complete light front wavefunction of the meson is then given as 
\begin{equation}
\Psi(\zeta, z, \phi)= e^{iL\phi} \mathcal{X}(z) \frac{\phi (\zeta)}{\sqrt{2 \pi \zeta}} \;,
\label{mesonwf}
\end{equation}
where the longitudinal wavefunction $\mathcal{X}(x)=\sqrt{z(1-z)}$, obtained by mapping the pion electromagnetic form factors in AdS and in physical spacetime \cite{Brodsky:2014yha}. For the vector mesons (like $\rho$, $K^*$ and $\phi$), we set $n=0, L=0$ and $S=1$  to obtain
\begin{equation}
\Psi_{0,0} (z,\zeta) = {\kappa \over \sqrt \pi}  \sqrt {z(1-z) }  \exp{\left[-{\kappa^2 \zeta^2 \over 2} \right] }\;.
\label{wavef}
\end{equation}
Allowing for small quark masses, the wavefunction becomes 
\beq  \Psi_{\lambda} (z,\zeta) = {\mathcal N}_{\lambda} \sqrt{z (1-z)}  \exp{ \left[ -{ \kappa^2 \zeta^2  \over 2} \right] }
\exp{ \left[ -{{(1-z)m_q^2+zm_{\bar q}^2} \over 2 \kappa^2 z(1-z) } \right]} \;.
\label{hwf}
\eeq
So far, the helicities of the quark and antiquark have been ignored. Assuming the spin structure of a vector meson is similar to that of a photon \cite{Forshaw:2012im}, we can restore the dependence of the light-front wavefunctions on the helicities of the quark and antiquark. The resulting twist-2 holographic DAs for the $K^*$ meson are then given by 
\begin{equation}
\phi_{K^*}^\parallel(z,\mu) =\frac{N_c}{\pi f_{K^*} M_{K^*}} \int \d
r \mu
J_1(\mu r) [M_{K^*}^2 z(1-z) + m_{\bar{q}} m_{s} -\nabla_r^2] \frac{\phi_{K^*}^L(r,z)}{z(1-z)} \;,
\label{phiparallel-phiL}
\end{equation}
\begin{equation}
\phi_{K^*}^\perp(z,\mu) =\frac{N_c }{\pi f_{K*}^{\perp}} \int \d
r \mu
J_1(\mu r) [m_s - z(m_s-m_{\bar{q}})] \frac{\phi_{K^*}^T(r,z)}{z(1-z)} \;,
\label{phiperp-phiT}
\end{equation}
where $f_{K^*}$ and $f_{K^*}^\perp$ are the vector and tensor couplings of $K^*$. The former can be accessed experimentally through the electronic decay width of the meson and this provides a first constraint for our DAs. In particular, it allows us to constrain the quark masses. 

\section{Results}
\begin{figure}[htbp]
	\begin{subfigure}{}
		\centering
		\includegraphics[width=0.3\textwidth]{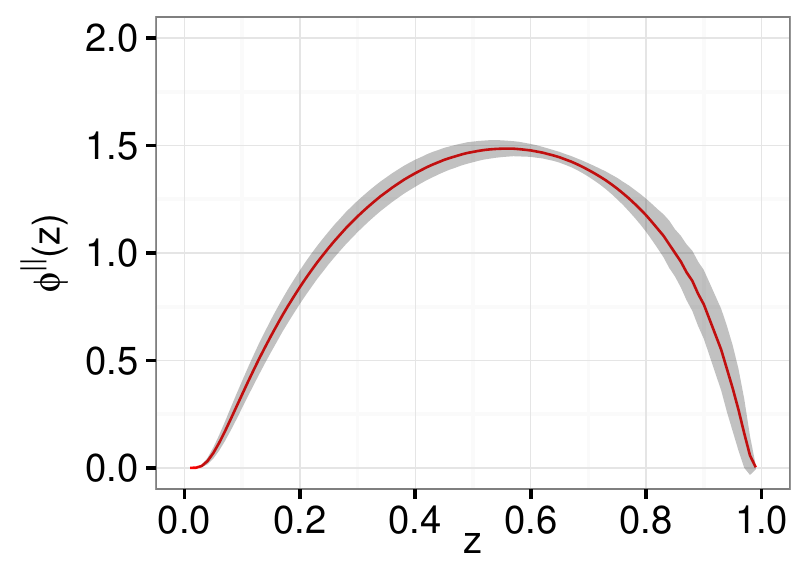}
		\label{adsdapara}
	\end{subfigure}
	\hspace{.1cm}
	\begin{subfigure}{}
		\centering
		\includegraphics[width=0.3\textwidth]{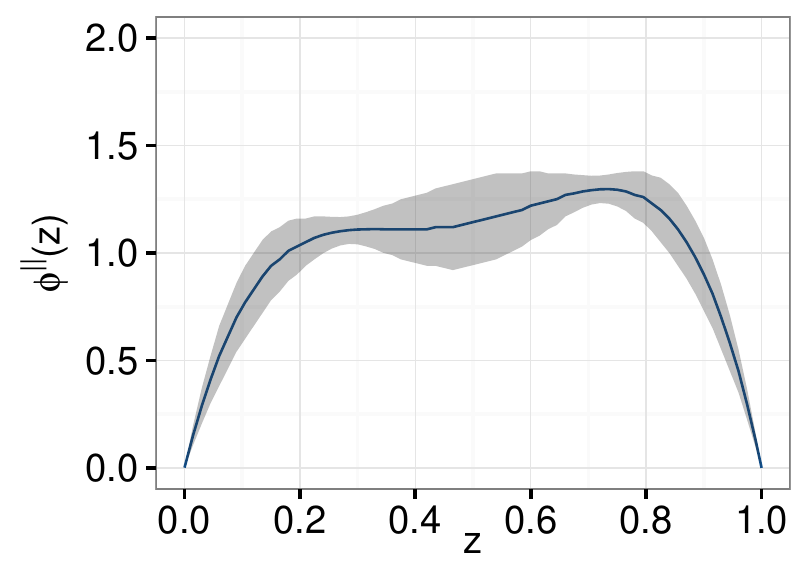}
		\label{srdapara}
	\end{subfigure}\\
	\begin{subfigure}{}
		\centering
		\includegraphics[width=0.3\textwidth]{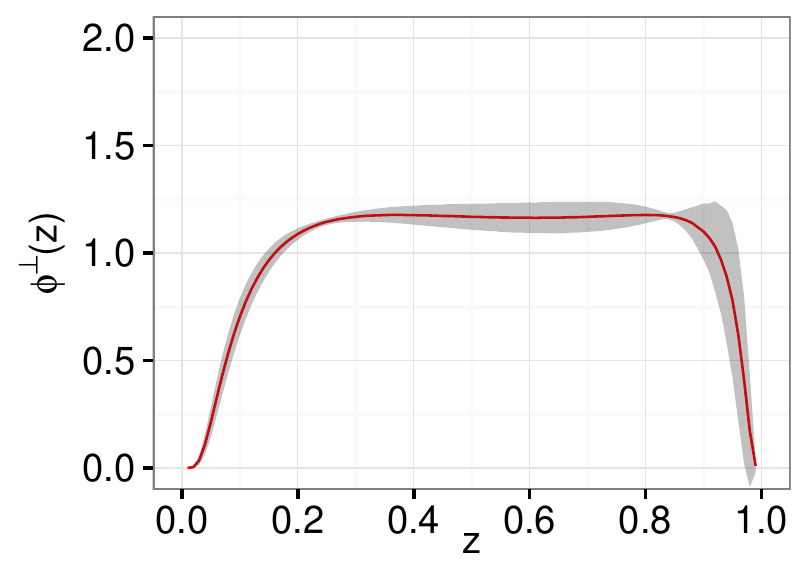}
		\label{adsdaperp}
	\end{subfigure}
	\begin{subfigure}{}
		\centering
		\includegraphics[width=0.3\textwidth]{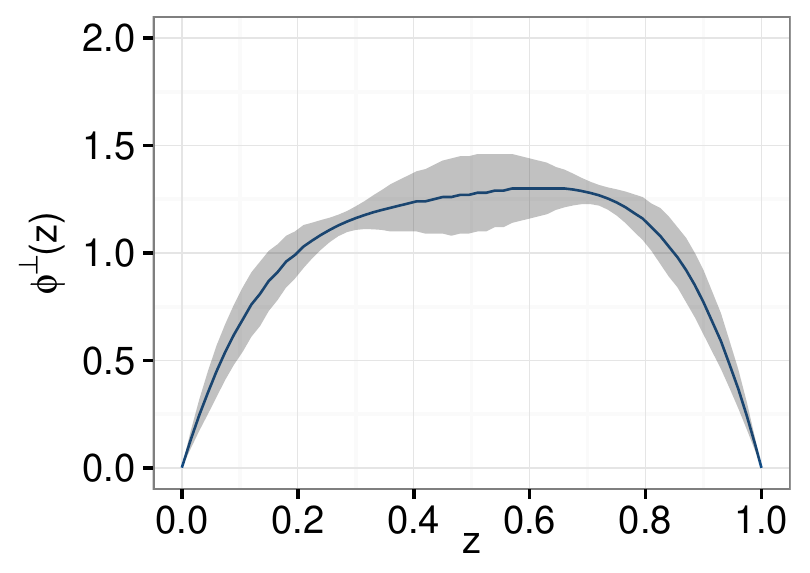}
		\label{srperp}
	\end{subfigure}
	\caption{Twist-2 DAs predicted by hQCD (graphs on the left) and SR (graphs on the right).  The uncertainty band is due to the variation of the quark masses for hQCD and the error bar on Gegenbauer coefficients for SR.}
	\label{das}
\end{figure}

Table \ref{tab:decay} shows the holographic QCD (hQCD) predictions for the $K^*$ vector and tensor coupling compared with the available experimental and lattice data.  We observe that quark masses $m_{\bar q}=195\pm 55$ MeV and $m_s=300\pm 20$ MeV leads to predictions consistent with data. 
\begin{table}[h]
	\[
	\begin{array}
	[c]{|c|c|c|c|c|c|c|}\hline
	\mbox{Approach}&\mbox{Scale}~ \mu  &m_{\bar{q}} \mbox{[MeV]} & m_s \mbox{[MeV]} &f_{K^*} \mbox{[MeV]} &f_ {K^*}^{\perp} (\mu) \mbox{[MeV]}&f_{K^*}^{\perp}/f_{K^*} (\mu)\\ \hline
	\mbox{AdS/QCD} & \sim 1~\mbox{GeV} & 140 & 280 & 200  & 118 & 0.59 \\ \hline
	\mbox{AdS/QCD} & \sim 1~\mbox{GeV} & 195 & 300 & 200  & 132 & 0.66 \\ \hline
	\mbox{AdS/QCD} & \sim 1~\mbox{GeV} & 250 & 320 & 200  & 142 & 0.71 \\ \hline \mbox{Experiment}  & &  & &205\pm 6  & & \\ \hline
	\mbox{Lattice}  & 2 ~\mbox{GeV}& & & & &0.780 \pm 0.008 \\ \hline
	\mbox{Lattice}  & 2 ~\mbox{GeV}& &  & &  & 0.74 \pm 0.02\\ \hline
	\end{array}
	\]
	\caption{Comparison between hQCD predictions for the decay constant of the $K^*$ meson with experiment (obtained from $\Gamma(\tau^- \to K^{*-} \nu_{\tau})$), and the ratio of couplings with lattice data.}
	\label{tab:decay}
\end{table}
In Fig. \ref{das}, we compare our predictions for twist-2 $K^*$ DAs with those obtained from QCDSR.  Indeed, the theoretical calculation of the isospin asymmetry in $B\to K*\gamma$ decay is directly sensitive to the end-point behaviour of the DAs and it turns out that the holographic DAs do not lead to the end-point divergences encountered with SR DAs. 
Using the above DAs within the light-cone sum rules (LCSR) method, we calculate $B\to K^*$ transition form factors (FFs) and in turn make predictions for observables like differential branching ratio and asymmetries. 
\begin{figure}[htbp]
	\begin{subfigure}{}
		\centering
		\includegraphics[width=0.4\textwidth]{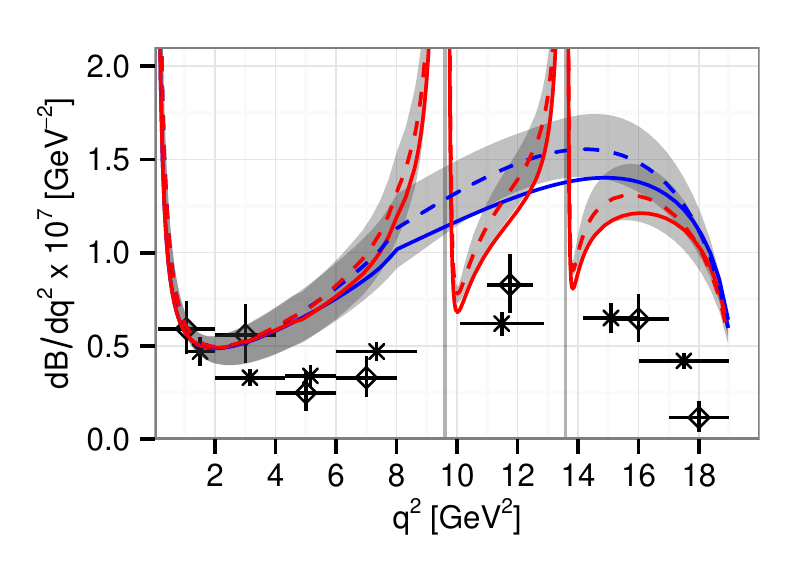}
		\label{adsdapara}
	\end{subfigure}
	\hspace{.01cm}
	\begin{subfigure}{}
		\centering
		\includegraphics[trim=0 0 0 10,clip, width=0.4\textwidth]{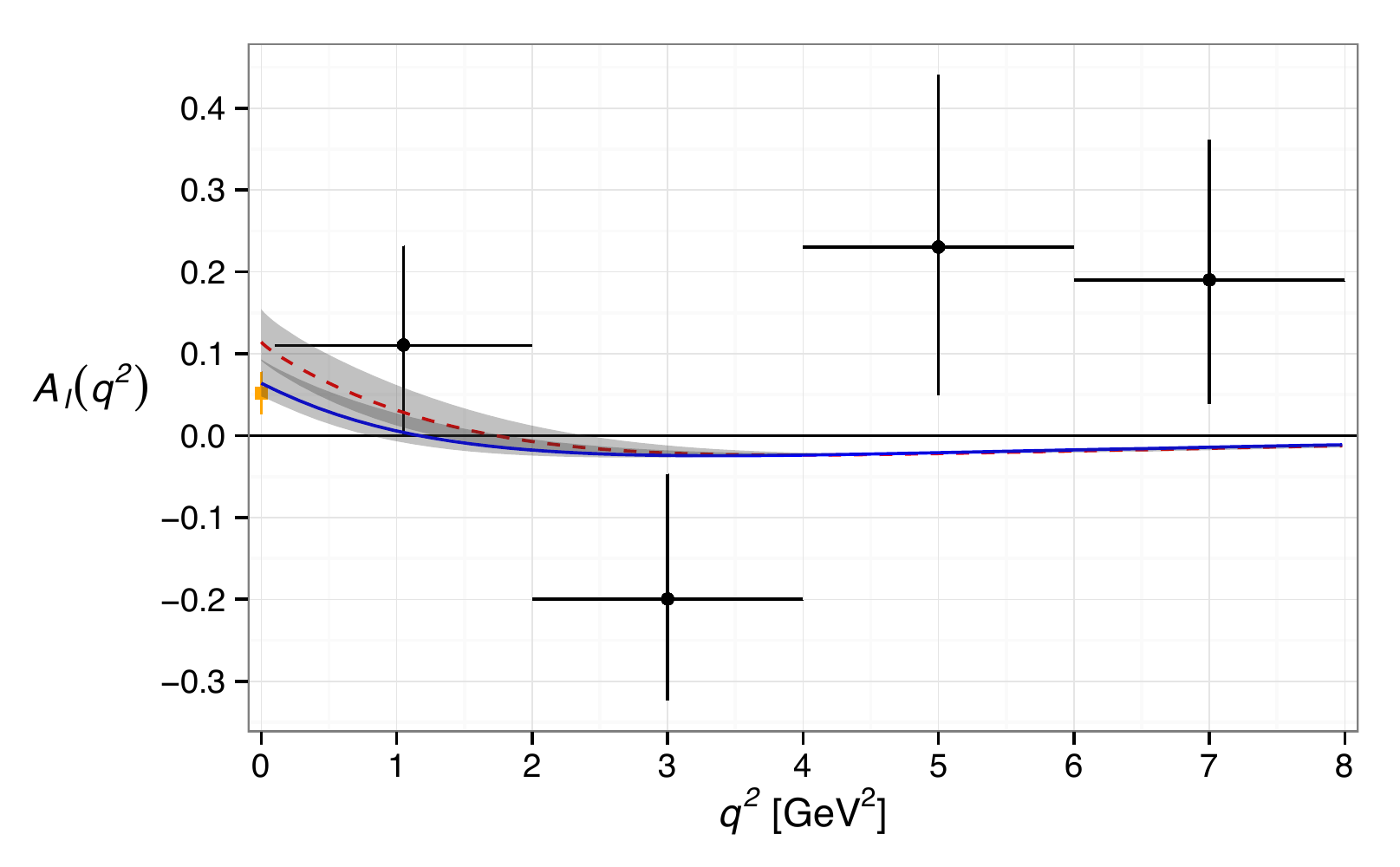}
		\label{srdapara}
	\end{subfigure}
	\caption{The differential branching ratio and isospin asymmetry for $B\to K^*\mu^+\mu^-$ predicted by hQCD (solid) and QCDSR (dashed) lines compared with the available experimental data.}
\end{figure} 
We observe that hQCD predictions are lower than those of the SR for the entire kinematical range of the momentum transfer ($q^2$).  We have also made predictions for the, yet to be measured, rare $B\to K^*\nu\bar{\nu}$ decay.
\begin{figure}
	\centering
	\includegraphics[width=.60\textwidth]{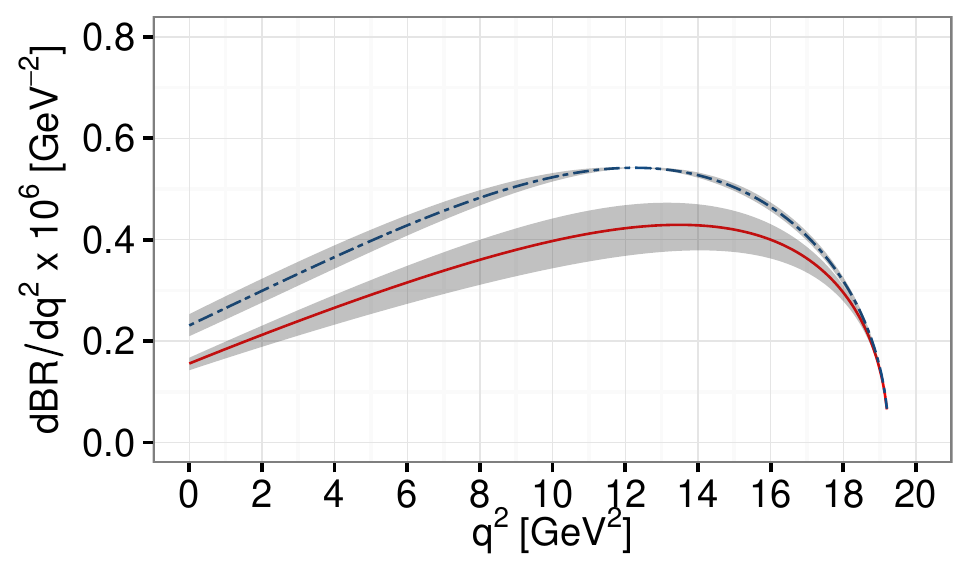} 
	\caption{The hQCD (Solid line) and QCDSR (Dashed line) predictions for the differential Branching Ratio for $B\to K^*\nu\bar\nu$.  The shaded band represents the uncertainty coming from the form factors.} 
	\label{fig:BKstarnu}
\end{figure}
In Figure \ref{fig:BKstarnu}, the differential branching ratio for $B\to K^*\nu\bar{\nu}$ as predicted by hQCD and QCDSR are presented.  This decay channel can provide a clean venue for testing these hadronic models.  It is interesting to note that the predictions significantly for low to intermediate $q^2$ range.

\section{Acknowledgment}	
MA would like to thank the organizers of FPCP 2019 for the wonderful experience.  This work is supported in part by MA and RS's NSERC grants.

\end{document}